\documentstyle[prl,aps,epsf]{revtex}
\begin{document}
\twocolumn[\hsize\textwidth\columnwidth\hsize \csname @twocolumnfalse\endcsname
\title{Temperature- and Bias-dependence of Magnetoresistance in Doped Manganite
Thin Film Trilayer Junctions}
\author{J. Z. Sun and D. W. Abraham}
\address{IBM T. J. Watson Research Center,
P. O. Box 218,
Yorktown Heights, NY 10598}
\author{ K. Roche and S. S. P. Parkin}
\address{IBM Almaden Research Center,
650 Harry Road,
Almaden CA 95120}
\date{\today }
\maketitle

\begin{abstract}
Thin film trilayer junction of La$%
_{0.67}$Sr$_{0.33}$MnO$_3$ - SrTiO$_3$ - La$_{0.67}$Sr$_{0.33}$MnO$_3$
shows a factor of 9.7 change in resistance, in a magnetic field around 100
Oe at 14K. The junction magnetoresistance is bias and temperature dependent. 
The energy scales associated with bias and temperature dependence are 
an order of magnitude apart. The same set of energies also determine the
bias and temperature dependence of the differential conductance of the 
junction. We discuss these results in terms of metallic cluster inclusions
at the junction-barrier interface.
\end{abstract}


\label{firstpage}
]
\narrowtext
\sloppy
%\newpage
Large low-field magnetoresistance (MR) has been \linebreak observed in trilayer thin
film junctions of La$_{0.67}$Sr$_{0.33}$MnO$_3$/SrTiO$_3$/La$_{0.67}$Sr$%
_{0.33}$MnO$_3$ \cite{142,224,293,191,390}. The mechanism is yet to be fully 
understood. These manganites are expected to be half-metals\cite{19,394} when
ferromagnetic order is fully developed. Their
low carrier concentration in the minority band makes its minority carrier
prone to disorder-induced localization\cite{394}. According to
spin-dependent tunneling models\cite{28,362,mr6,299}, a  half-metallic 
metal-insulator-metal junction would exhibit large, almost 
infinite MR. However, the observed bias-
and temperature-dependence don't correspond to
a clean metal-insulator-metal tunneling process. The junction resistance
varies strongly with temperature, especially above $130K$, indicating 
a barrier with high concentration of
defects\cite{224,191,390}. 
The MR disappears prematurely above $130K$, well
below the Curie temperature $T_c=360K\,$of the electrodes. In addition, the
junction MR is strongly bias-dependent when a bias-voltage of
the order of $0.1V$ is applied.

Here we quantify the temperature- and bias-dependent
magnetotransport properties of junctions made of La$_{0.67}$Sr$_{0.33}$MnO$%
_3 $/SrTiO$_3$/La$_{0.67}$Sr$_{0.33}$MnO$_3$ (LSMO/STO/LSMO) and La$_{0.67}$%
Sr$_{0.33}$MnO$_3$/Al$_2$O$_3$ \linebreak/Permalloy (LSMO/AO/Py) trilayers. 
A low-bias minimum develops in the differential conductance for
temperatures below $120K$, indicating the possible presence of a
Coulomb gap. Two energy scales are present: first, an
energy corresponding to the bias voltage for the suppression of MR.
This energy happens to coincide with the bias level for the onset of a
low-bias conductance minimum; secondly, an energy associated with the
high-temperature suppression of MR, which happens to be identical to
the temperature for the disappearance of the low-bias conductance
minimum. We also show LSMO/STO/LSMO junctions with an order of
magnitude change in resistance in 100 Oe, indicating that the 
intrinsic MR in these junctions is likely to be very large.

The fabrication process for LSMO/STO/LSMO junctions have been discussed in
detail before\cite{142,136,157}. The process for LSMO/AO/Py is similar. In
both cases the base LSMO film is around $500\sim 600\AA $ thick. The top
LSMO film is around 400\AA . The LSMO\ films are grown epitaxially on LaAlO%
$_3$(100) or on NdGaO$_3$(110) substrates using laser ablation.

\begin{figure}[tbp]
%\epsfxsize=3in
%\epsfbox{fig1.ps}
\caption{An AFM image of a representative film surface, showing a
peak-to-peak roughness around 15\AA . An accompanying scanning electron
micrograph is shown to the right for comparison.}
\label{Fig1}
\end{figure}

The surface of these laser ablation-grown LSMO films are quite smooth. An
example is shown in Fig.\ref{Fig1}, where an atomic force microscopy (AFM)
image was taken on a single layer LSMO film, 1000\AA\ thick, grown using
the above mentioned deposition process on NdGaO$_3$ (110) surface. The
peak-to-peak surface roughness in this case is around 15\AA , excluding
laser particulates which has a typical density of around $\sim 10^6/cm^2$.

For LSMO/STO/LSMO junctions, the entire trilayer structure was made {\it in
situ}. For LSMO/AO/Py trilayers, the film samples were transported to a
different vacuum system after the deposition of LSMO. 
The wafer was first cleaned by exposure to $30sec$ of oxygen plasma. The Al$_2$O$_3$
barrier layer was formed by sputter deposition of 12\AA\ of aluminum,
followed by $120sec$ plasma oxidation\cite{222}. The
Permalloy counter electrode was then sputter deposited, 120\AA\ thick, in a 
process similar
to that used in metallic magnetic tunneling junctions\cite{222}. For both
type of films the same photolithography process was used to define the
current-perpendicular junction structure.

\begin{figure}[tbp]
\epsfxsize=3in
\epsfbox{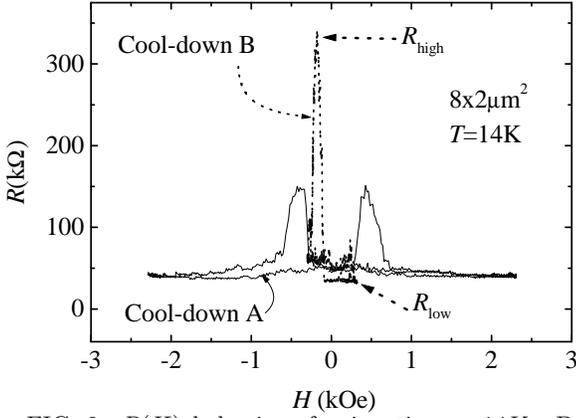}
\caption{$R(H)$ behavior of a junction at $14K$. Data are 10-trace
averaged. Field was swept at $0.13Hz$. Junction bias
was $10nA$. After cool-down A, a factor of $2$ change in $R$ was seen. 
A very large MR, of $R_{high}/R_{low}=9.7$ was seen at low
sweeping amplitude of around $100Oe$ during another cool-down(B).}
\label{Fig2}
\end{figure}

An example of the dc junction resistance vs. applied field, $R(H)$, is shown
in Fig.\ref{Fig2}. The junction pillar was $8\mu m\times 2\mu m$ in size,
sitting on a $40\mu m$ wide base stripe, with its long axis perpendicular to
the direction of the stripe. The magnetic field was applied
parallel to the film surface along the long axis. The measurement
temperature was 14K, and the junction was biased at $10nA$. From $R(H)$
loops as these one defines a junction resistance $R_{high}$, corresponding
to the resistive-high state of the junction, and a $R_{low}$ for the
resistive-low state.

The $R(H)$ behavior is history dependent. After cool-down (A), a factor
of $2$ change in $R$ was seen on the 10-trace averaged $R(H)$.
During another cool-down (cool-down (B) in Fig.\ref{Fig2}), when the field
sweep amplitude was first opened up from zero, a very large MR, of $%
R_{high}/R_{low}=9.7$, was seen at low sweeping amplitude of around $100Oe$.
These results highlight the importance the electrode's magnetic state has
on junction transport properties. It is likely that the intrinsic MR in
these structures is even larger, and that the actual MR one observed is
still limited by the distribution of magnetic domains.

\begin{figure}[tbp]
\epsfxsize=3in
\epsfbox{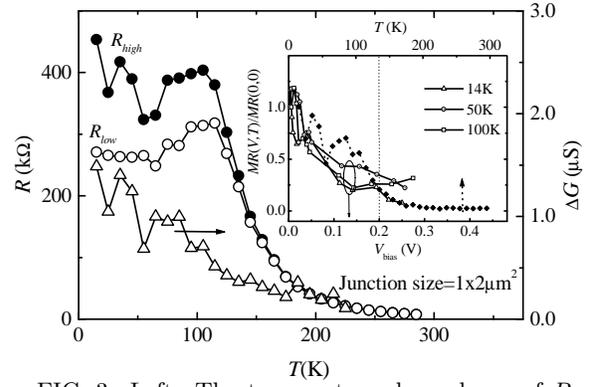}
\caption{Left: The temperature dependence of $R_{high}$ and $R_{low}$. 
Right: conductance difference between $R_{high}$ and $R_{low}$ showing 
continuous decrease for $T<130K$. Inset: Bottom axis, bias-dependence 
of the normalized $MR$ (i.e, $MR(V,T)/MR(0,0)$) for three different 
temperatures. Although the magnitude of $MR$ varies
with temperature, its bias-dependence remained essentially temperature
independent. Top axis: normalized $MR$ vs. temperature.}
\label{Fig3}
\end{figure}

Fig.\ref{Fig3} shows the temperature dependence of $R_{high}$ and $R_{low}$. 
Both $R_{high}$ and $R_{low}$ are taken
from 10-trace averaged $R(H)$ loops. For $T>130K$, they follow an $\exp
\left[ \left( T_o/T\right) ^{1/4}\right] $ scaling \cite{293}. Below $130K$, 
$R_{high}$ and $R_{low}$ branch apart, giving an MR that increases upon decreasing
temperature. The noise in data is again a consequence of magnetic
instabilities of the electrodes.

Also shown in Fig.\ref{Fig3} is the conductance difference 
$\Delta G=\frac 1{R_{low}}-\frac 1{R_{high}}$ vs.
temperature. A continuous decrease was seen of $%
\Delta G(T)$ at the low temperature end, where $R_{high}$ and $R_{low}$ are
only weakly temperature dependent. This rules out the parallel shunt
mechanism\cite{293} as an explanation for the decrease of MR in this temperature
region, because parallel shunt from MR-inactive channels
would have a constant $\Delta G(T)$.

The junction MR is bias-dependent. The inset of Fig.\ref{Fig3} shows the
bias-dependence of junction MR at several temperatures. The $MR$ is
suppressed to $\sim 25\%$ of its low-bias value at a bias-voltage of $0.2V$.
Although the magnitude of $MR$ decreases rapidly with increasing $T$, the
characteristic bias-dependence of which remained essentially temperature
independent. Shown in the same plot is $MR$ as a function of temperature.
Here $MR=\frac{R_{high}-R_{low}}{R_{low}}$, all resistances are dc.

Data in Fig.\ref{Fig3} inset reveal two energy scales. The first relates to the
bias-voltage: $E_b\sim 200meV$. This corresponds to the bias level at which
MR is suppressed. The second relates to the temperature: $E_T\sim 130K=11meV$%
, corresponding to the temperature that suppresses MR. The two energy scales
are an order of magnitude apart.

\begin{figure}[tbp]
\epsfxsize=3in
\epsfbox{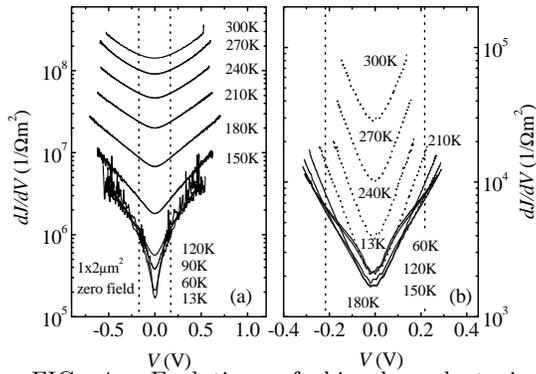}
\caption{Evolution of bias-dependent junction differential conductance
with temperature. (a)For an LSMO/STO/LSMO junction.
The two energy scales, $E_{b}$ and $E_{T}$ are again present. (b)Same 
measurement for an LSMO/AO/PyO junction, showing the same energy scales.}
\label{Fig4}
\end{figure}

Fig.\ref{Fig4}(a) shows the evolution of the
differential junction conductance $\sigma _D=\frac{dJ}{dV}$ with
temperature. The high temperature and high bias-voltage part of $\sigma
_{D}\,$probably represents conduction via shallow defect states in 
the barrier\cite{3} which we are not going to discuss. Instead, notice that for
temperatures below $120K$ ($\sim E_T$), a conductance minimum develops for
bias-voltages below $200meV$ ($\sim E_b$). Again, $E_b$ and $E_T$ show up
as the energy scales associated with bias and temperature dependence.

The development of a low-bias conductance minimum also occurs in LSMO/AO/Py
junctions, as shown in Fig.\ref{Fig4}(b). Again, there is a characteristic
temperature of $T=180K$ close to $E_T\,$observed before, and a
characteristic voltage of $200meV$ for low-bias conductance minimum, similar
to the $E_b$ of data in Fig.\ref{Fig4}(a).

The fact that both $E_T\,$ and $E_b$ are present for junctions made with
different barrier and counter-electrodes suggest the mechanism(s)
responsible for them are associated with the base electrode LSMO film,
possibly with its interface at the barrier, rather than with the specifics
of the barrier physics.

One mechanism that could
produce conductance minimums over a wide range of bias conditions is
metallic inclusions at the junction interface. The metallic inclusions, if
small enough, will show a Coulomb gap in its low-bias conductance \cite{439}%
. The conductance minimum in this case relates to the effective capacitance
of the inclusion cluster, which is proportional to the cluster
size, roughly speaking. Assuming the charging energy can be estimated
as\cite{249,404,405} 
$E_c=\frac{e^2}{2C}=\frac 1{8\pi \varepsilon \varepsilon _o}\left( \frac{e^2}d%
\right) F_o $,
where $e$ is electron charge, $C$ is the capacitance of the cluster, $%
\varepsilon $ the matrix dielectric constant, $d$ the diameter of the
cluster, $F_o$ is a form factor that depends on the details of the local
environment of the cluster and its distance to other conducting structures ($%
F_o=2$ for an isolated cluster), and assume $\varepsilon =5$, $F_o=2$ (See 
\cite{404} for estimate of $F_o$), 
one gets $d\approx 15\AA $. 
Is it possible that there are metallic clusters at the LSMO\ surface of 
dimensions
around $15\AA $? Could these be related to the localization length of the
spin-polarized carriers? these are open questions at present.

The conductance minimum from a single Coulomb gap is very sharp at low bias\cite{439}.
A distribution of $E_c $ is likely to be present that could smear out the sharp 
low-bias cusp in the Coulomb gap's conductance, making it appear similar to
the shape observed.

The hypothesis of a Coulomb gap explains the bias dependent behavior, but it
would not explain the temperature dependence. Since $E_T\,$ is an order
magnitude lower than $E_b$, some other mechanism(s) must be involved in
suppressing the gap formation process for temperatures above $E_T$.

We wish to thank John
Slonczewski, Bill Gallagher, Roger Koch, Arunava Gupta, Steve Brown, 
John Connolly at IBM Research Yorktown Heights; 
Xinwei Li, Yu Lu and Gang Xiao from Brown
University for helpful discussions and for assistance at various stages of
the experiment.

%\bibliographystyle{prsty}
%\bibliography{dump,all98dmp,mrdump}

\begin{references}

\bibitem{142}
J.~Z. Sun, W.~J. Gallagher, P.~R. Duncombe, L. Krusin-Elbaum, R.~A. Altman, A.
  Gupta, Y. Lu, G.~Q. Gong, and G. Xiao, Appl. Phys. Lett. {\bf 69},  3266
  (1996).

\bibitem{224}
J.~Z. Sun, L. Krusin-Elbaum, A. Gupta, G. Xiao, P.~R. Duncombe, and S.~S.~P.
  Parkin, IBM J. of Res. and Dev. {\bf 42},  89  (1998).

\bibitem{293}
J.~Z. Sun, L. Krusin-Elbaum, P.~R. Duncombe, A. Gupta, and R.~B. Laibowitz,
  Appl. Phys. Lett. {\bf 70},  1769  (1997).

\bibitem{191}
Y. Lu, X.~W. Li, G.~Q. Gong, G. Xiao, A. Gupta, P. Lecoeur, J.~Z. Sun, Y.~Y.
  Wang, and V.~P. Dravid, Phys. Rev. B {\bf 54},  R8357  (1996).

\bibitem{390}
M. Viret, M. Drouet, J. Nassar, J.~P. Contour, C. Fermon, and A. Fert,
  Europhysics Letters {\bf 39},  545  (1997).

\bibitem{19}
W.~E. Pickett and D.~J. Singh, Phys. Rev. B {\bf 53},  1146  (1996).

\bibitem{394}
W.~E. Pickett and D.~J. Singh, J. Magn. and Magn. Mat. {\bf 172},  237  (1997).

\bibitem{28}
J. Slonczewski, IBM Technical Disclosure Bulletin {\bf 19},  2328  (1976).

\bibitem{362}
M. Julliere, Phys. Lett. {\bf 54A},  225  (1975).

\bibitem{mr6}
J.~C. Slonczewski, Phys. Rev. B {\bf 39},  6995  (1989).

\bibitem{299}
A.~M. Bratkovsky, Phys. Rev. B {\bf 56},  2344  (1997).

\bibitem{136}
J.~Z. Sun, L. Krusin-Elbaum, S.~S.~P. Parkin, and G. Xiao, Appl. Phys. Lett.
  {\bf 67},  2726  (1995).

\bibitem{157}
J.~Z. Sun, L. Krusin-Elbaum, A. Gupta, G. Xiao, and S.~S.~P. Parkin, Appl.
  Phys. Lett. {\bf 69},  1002  (1996).

\bibitem{222}
W.~J. Gallagher, S.~S.~P. Parkin, Y. Lu, X.~P. Bian, A. Marley, R.~A. Altman,
  S.~A. Rishton, K.~P. Roche, C. Jahnes, T.~M. Shaw, and G. Xiao, J. Appl.
  Phys. {\bf 81}, 3741 (1996).

\bibitem{3}
Y. Xu, D. Ephron, and M.~R. Beasley, Phys. Rev. B {\bf 52},  2843  (1995).

\bibitem{439}
H.~R. Zeller and I. Giaever, Phys. Rev. {\bf 181},  789  (1969).

\bibitem{249}
J.~S. Helman and B. Abeles, Phys. Rev. Lett. {\bf 37},  1429  (1976).

\bibitem{404}
P. Sheng, B. Abeles, and Y. Arie, Phys. Rev. Lett. {\bf 31},  44  (1973).

\bibitem{405}
P. Sheng, Phil. Mag. B {\bf 65},  357  (1992).

\end{references}

\end{document}